# CMOS-based Single-Cycle In-Memory XOR/XNOR

Shamiul Alam[1], Jack hutchins[1], Nikhil Shukla[2], Kazi Asifuzzaman[3], and Ahmedullah Aziz[1]*

[1]Dept. of Electrical Eng. and Computer Sci., University of Tennessee, Knoxville, TN, 37996, USA.
[2]Department of Electrical and Computer Engineering, University of Virginia, Charlottesville, VA, USA.
[3]Oak Ridge National Laboratory, Oak Ridge, TN, USA.

*Corresponding Author. Email: aziz@utk.edu

*Abstract-* **Big data applications are on the rise, and so is the number of data centers. The ever-increasing massive data pool needs to be periodically backed up in a secure environment. Moreover, a massive amount of securely backed-up data is required for training binary convolutional neural networks for image classification. XOR and XNOR operations are essential for large-scale data copy verification, encryption, and classification algorithms. The disproportionate speed of existing compute and memory units makes the von Neumann architecture inefficient to perform these Boolean operations. Compute-in-memory (CiM) has proved to be an optimum approach for such bulk computations. The existing CiM-based XOR/XNOR techniques either require multiple cycles for computing or add to the complexity of the fabrication process. Here, we propose a CMOS-based hardware topology for single-cycle in-memory XOR/XNOR operations. Our design provides at least 2× improvement in the latency compared with other existing CMOS-compatible solutions. We verify the proposed system through circuit/system-level simulations and evaluate its robustness using a 5000-point Monte Carlo variation analysis. This all-CMOS design paves the way for practical implementation of CiM XOR/XNOR at scaled technology nodes.**

## I. Introduction

Academia and industry are pushing their last strides in keeping Moore's law alive, demonstrated by IBM's 2 nm process technology [1]. However, as the available bandwidth between the processor and main memory is not growing commensurately with the advancements in compute units, the well-known 'memory wall' [2] is becoming one of the toughest challenges for engineers in this exascale (big data) computing era. The issue with handling this massive data load is getting more acute with unprecedented progress in machine learning and artificial intelligence (AI) applications. Surprisingly, these data-intensive applications are often not inherently complicated. Rather, they rely on simple logic operations at a massive scale. Therefore, data movement ends up being the bottleneck, causing latency issues and consuming more energy than the computation process itself [3]. Recent reports by *Google* have shown that a significant portion of their data center workload is performing bulk data movement and about 20-42% of the energy is required to drive the data bus connecting the compute and memory units [4], [5]. This specific data movement problem is causing the traditional *von Neumann* architecture to lose its glory, where back-and-forth data movement is necessary between the memory and compute units. As an alternative, compute-in-memory (CiM) has garnered attention in the research community [2], [6]–[10]. CiM not only dramatically reduces the data movements, but also takes advantage of large internal memory bandwidth and enables massive parallelism to improve latency. In addition to the endeavor to improve the architecture, device engineers are exploring next-generation memory technologies as the mainstream CMOS memories are approaching the scaling limit [11]–[15]. The emerging memories are expected to provide a faster yet more energy-efficient solution in a compact footprint. Combining the best of both worlds, several CiM architectures have been proposed in recent years with emerging memory devices [8], [16]–[22]. However, with exponentially increasing data volume, customized solutions are needed for optimized performance in application-specific scenarios.

A denser integration of memory chips with CiM capability will radically change the data center field. With the advent of cloud computing, consumer computer applications are gradually finding their way into virtual machines rather than physical devices, thereby leading to more data in data centers. Keeping this ever-increasing data in a secured backup is a challenging task in terms of performance, energy, and memory. While intelligent and efficient algorithms were proposed for bulk data movement in data centers using row-level cloning [23], integrity verification of the copy procedure is also extremely important. Moreover, in the age of cybersecurity and identity theft, data encryption is equally crucial. Having such securely backed-up data is essential for big data applications like image classification. This data can be used to train the classifier algorithms to get acceptable inference accuracy. Fig. 1 illustrates the usage of XOR/XNOR operations for copy verification, encryption, and image classification algorithms in binary convolutional neural networks (CNN).

Here, we propose a ubiquitous system to achieve single-cycle in-memory bitwise XOR/XNOR operation using modified peripheral sensing circuitry. We evaluate the functionality and robustness of our design using transient simulations in HSPICE, and *Monte Carlo* variation analysis. While our proposed design is aimed at fast and secure data movement at the data center level, it can also be used in the most prominent big-data application for AI- a 'binary CNN hardware accelerator' with no additional operation cycle. We discuss the motivation and principle of in-memory XOR/XNOR in section II. We then present our design methodology and the simulation framework (section III). Sections IV and V present the timing simulations and variation analysis, respectively. Section VI presents a comparison with existing literature.

## II. Motivation for Single-Cycle In-Memory XOR/XNOR

Bulk data copy is such an expensive process (in terms of memory usage and energy demands), that there has been a separate hardware-level instruction set for it since the introduction of Intel IA-32 architecture [24]. Extensive studies have shown that the optimized way to transfer data between the conventional memory cells is at the row granularity [23], [25]. In cutting-edge memory chips, an entire row of data is copied from the memory array to the corresponding row buffer and then to the destination row [26]. This multi-cycle copy procedure is already a major concern for low-latency memories. On top of that, the need for additional cycles to validate a successful copy operation aggravates the issue.

For the validation process, parity checking is the most commonly used algorithm in present-day digital electronics. An odd parity checker performs XOR operations between the bits copied from and to the memory cells. A logical '0' XOR output indicates a successful copy operation (Fig. 1(a)). In addition to having back-ups, it is also important to ensure its security. Fortunately, the in-memory XOR operation is perfectly suited for data encryption (Fig. 1(b)). Among the known techniques for ciphers, XOR is the most trustworthy and unbreakable if the key used is a true random number.

The significance of performing such XOR/XNOR operations within the memory block (CiM implementation) can be well understood with a system-level view. Two subsequent row activation cycles are needed to copy a single unique row of data into the memory block. At least another cycle is required for verification of each 'copy' incidence using XOR operation. For a memory bank of 512 rows, duplication of 256 unique data rows requires $256 \times 2$ row activation cycles and 256 XOR operation streams for verification (assuming single-cycle XOR).

Similarly, every single row goes through a single XOR operation stream with a key stored in the row buffer for encryption. Now, if each of these XOR operations is itself a multi-cycle process, the latency will take a serious hit. Now, all the in-memory XOR operations previously demonstrated take more than one cycle except for one proposed in [17], which too is a memristor-only CMOS non-compatible design, for which the design space will be too complicated. To the best of our knowledge, ours is the first CMOS-compatible in-memory XOR that operates in a single-cycle. We propose a simple all-CMOS-based peripheral circuit design, slightly modifying the sensing circuitry to employ CiM XOR for superior performance in bulk data operations. On top of that, this modification in peripheral circuitry can also be used in binary neural networks like image classification problems, which is essentially an XNOR operation (shown in Fig. 1(c)). Thus, to gain excellent capacity and speed in an in-memory system, the proposed system can be put into use.

### III. Design Methodology and Simulation Framework

For a conventional memory array comprised of access transistors and memory cells, the sense line (SL) currents are collected and sensed via a current-based sense amplifier at the periphery (Fig. 2(a)). In our work, we utilize the current-based sense amplifier (CSA) reported in [27] as the building block for the modified peripheral circuitry to realize the in-memory XOR/XNOR. Here, we use a ReRAM as the NVM cell, but the peripheral circuit modification (all CMOS) to realize the in-memory XOR/XNOR operation is a memory-agnostic design. Irrespective of the memory used, when in computation mode, two-word lines (WL) are asserted in a single sensing line to select the memory cells that will undergo the XOR/XNOR operation. The current contribution of the two selected cells along with the unselected ones of that column is fed into the modified SA. The modified SA consists of a current mirror to copy the SL current, two current-based SAs (CSA), one inverter, and one AND gate as shown in Fig, 2(c). Fig. 2(d) shows the circuit schematic of each CSA used in the SA. The SL current being fed into the two CSAs sets a gate voltage through the current mirror circuit. This set voltages then being compared to the reference voltages, produce binary outputs. As for XOR/XNOR operations, two different reference current levels are being used, they will produce two different logic outputs. These two different logic outputs, one negated through an inverter and the other one intact, fed into the AND gate, give out XOR/XNOR logic. Here it is noteworthy that, the complementary reference current level is set for two CSAs for giving out XOR/XNOR logic output. A truth table for the sense line current levels and the corresponding logic levels are shown in terms of the reference level set in Fig. 2(b). It can be seen from the illustration that reference current levels are set in between the $I_{00}$ and $I_{11}$ current levels. The reference currents are set in such an intelligent way that an AND operation of the outputs of two CSAs gives out the desired XOR/XNOR result. The sense amplifiers being exactly similar in construction in a CMOS process separates the two extreme cases of both the selected cells storing either '0' or '1' using two reference currents ( $I_{00} < I_{REF1} < I_{01}$ & $I_{01} < I_{REF2} < I_{11}$ ). This slight modification in peripheral sensing circuitry allows normal memory mode operation as well as single cycle XOR/XNOR operation, which can be very crucial in certain specific application scenarios. Not only that, but this design can also be used to implement other logic operations like AND/NAND, OR/NOR, etc. by carefully choosing the two reference current levels.

In this work, a rigorous SPICE simulation is done for the CiM provision in the memory array. For simulation, a phenomenological compact model of resistive RAM (ReRAM) is used as the non-volatile memory (NVM). The model is calibrated and matched with the experimental data for the Cu/HfO2/Pt stack published in [28]. The low resistance state (LRS) and the high resistance state (HRS) are set at 10 kΩ and 3 GΩ, respectively. 14 nm PTM (Predictive Technology Model) [29] transistors are utilized to simulate the CMOS transistors (FinFETs) used in the memory array and peripheral circuitry. A detailed Monte-Carlo variation analysis is also shown to determine the limitation of the effect of variation on the number of allowed rows in the memory array along with sense margins for the successful operation.

### IV. Functional Verification

Upon setting up the simulation framework, functional verification was performed for the in-memory XOR/XNOR operation in HSPICE. The memory array functions as expected in the memory mode, allowing successful write operations shown in Fig. 3. In the memory mode of operation, the bit lines (BL) are kept precharged and the access transistors are turned on for the selected cell applying suitable biases to the WLs and SLs. Although all the bit lines are kept high, the access transistors are not turned on for the unaccessed and half-accessed cells. Depending on the bias voltages applied to the WLs, BLs, and SLs, the corresponding memory state is stored in the memory cell. 0.4 V (-0.15 V) is applied to the corresponding BL for writing '1' ('0') into the memory cell, as per the non-volatile memory material we are using from [28]. Later, when WLs are asserted, the accessed cell gets the write voltage applied to the BL. The biasing scheme for write operations is designed in such a way that the half-accessed and unaccessed cells are not accidentally disturbed. Also, reading from the memory cell, we propose to use the same SA designed for the in-memory XOR/XNOR operation to make the peripheral circuitry universal for both memory and compute mode. The only difference between memory read and compute operation is that for reading, only one cell is accessed at a time, and reference current levels are different.

The proposed design allows the memory operations (write and read)as well as the in-memory logic operations. In the computation mode, the memory states stored in the accessed cells are first read and then the logic operation is done using the peripheral circuits. To demonstrate the successful operation of our design, we simulated a 3x3 array shown in Fig. 4(a). Here, all the bit lines (BL) are pre-charged with a 100 mV supply. After the WLs corresponding to two computing rows are asserted, current starts to flow through the memory cells. Fig. 4(b) shows the biasing scheme for the in-memory operations. Now, based on the assumed memory states for the accessed cells (shown in Fig. 4(a)), different current levels are obtained in the SLs. The SL current levels for different combinations of memory states in the columns are well distinguishable as shown in Fig. 4(d). Considering the unaccessed cells in HRS, the SL currents are obtained as 100 pA, 7.87 $\mu$A, and 15.7 $\mu$A for '00', '01'/'10', and '11' logic combinations in the accessed cells, respectively. The reference current levels of the sense amplifiers need to be carefully set in based on these numbers.

For verifying the XOR operation, we set the reference currents as $I_{REF1}$= 4 $\mu$A and $I_{REF2}$= 12 $\mu$A. When the SEN (Sense Enable) is enabled, the CSAs sense the current levels and result in logic '1' or '0' based on the difference between the SL currents and the reference currents. With the AND operation as shown in Fig. 2(b), the output of the XOR operation is obtained as shown in Fig. 3(d). As seen, the XOR-OUT becomes logic '1' only for '01'/'10' logic combination. Note, the SL currents are readily available in the sense amplifiers when WLs and BLs are asserted. Therefore, the XOR operation only requires a single cycle, for the AND operation to be completed. However, for XNOR operation, the reference currents are set in the exact opposite fashion ($I_{REF1}$= 12 $\mu$A and $I_{REF2}$= 4 $\mu$A). XNOR operation also requires a single-cycle.

## V. Variation Analysis

It is seen in Fig. 4(d) that the SL currents are well-distinguishable for different memory combinations in the cells in a single column. However, a quantitative analysis was performed to full-proof the robustness of the design. Even when a cell is not accessed (WL not asserted), a small leakage current flows through those cells: 28 pA for HRS and 774 pA for LRS. The leakage currents through the unaccessed cells contribute to the SL current of the column, which causes a risk of identifying the SL current of one logic combination as another. Therefore, the leakage current (depending on the LRS and HRS values) puts a restriction on the maximum number of rows allowed in an array. Also, average power consumption and area are two very important parameters that directly affect the scaling of the memory system. Fig. 5(a) and 5(b) show the effects of a number of fins on the power consumption and area of the CSA and the effects of HRS and LRS values on the maximum number of rows in the array, respectively. In Fig. 5(b), we show the effects of variation in both HRS and LRS separately which shows that the variation in LRS affects more significantly compared to that in HRS. With a fixed HRS, when we vary the LRS by changing the HRS/LRS ratio (black line in Fig. 5(b)), we observe that a larger HRS/LRS ratio results in higher scalability. This analysis not only lets a designer be aware of the size limitation of the memory array but also opens up a new window of research from the perspective of the material choice.

Furthermore, a rigorous 5000-point Monte-Carlo simulation is performed to ensure that different current levels are well-distinguishable even with the process variations. In our variation analysis, we consider a Gaussian distribution for LRS and HRS with a mean value of 10 kΩ and 3 GΩ (respectively) and a $3\sigma$ variation of 10% of the mean value. We also consider a variation in the threshold voltages of the transistors with a standard deviation of 25 mV. The results are shown in Fig. 5(c) and 5(d). Fig. 5(e) shows the schematic of a conventional current sense amplifier with different important nodes marked. The distribution in SL currents shown in Fig. 5(b) leads to a distribution of voltages at the $n_{CELL}$ node of the sense amplifier. Finally, the digital output at the OUT node is obtained based on the difference between the voltages set at $n_{CELL}$ and $n_{REF}$ nodes.

## VI.    Comparative Study

The surge in compute-in-memory research because of the 'memory wall' problem led to many recent publications. Studies have shown that the ReRAM crossbar array can implement logic operations in the crossbar array [30]. However, some of them are not necessarily fitted to the CiM concept as they use the memory technique to implement processing units. They still pay for the expensive data fetching from the memory and are limited by the memory bus bandwidth. Those that implement the in-memory computation, are tailored to do basic logic operations like AND, OR, etc., some to make ADD operations. Our work can be distinguished from those works in terms of bulk data application in an all-CMOS process.

Based on the required operation steps and overhead circuitry, a comparison with the existing relevant works [17], [30]–[33] is presented in Table I. Our work promises the most efficient solution in terms of latency. Also, an all-CMOS design makes it easy to implement.

We also extend the comparison to the application level using XNOR-Net which uses XNOR operation to replace the computationally complex convolution operations in convolutional neural networks (CNN). XNOR-Net is a CNN that uses binary filters and XNOR operations to decrease memory cost and decrease computational cost by around 58× [34]. Fig. 6(a) shows a single convolutional block of XNOR-Net. In the beginning, XNOR-Net performs batch normalization and then performs binary activation that binarizes the inputs and generates the scaling factors $K$ and $a$. From there, the XNOR convolution is performed. We propose using our XNOR processor to accelerate this part of the network. After calculating the XNOR convolution, we then perform element-wise multiplication with the scaling factors ($K$ and $a$) that we calculated before the XNOR operations. While these operations must be done outside of our accelerator, there are far fewer of these operations than XNOR operations, making our approach still viable despite this limitation. The theoretical speedup due to the use of XNOR convolution is given by [34]-

$$S = \frac{cN_W N_I}{\frac{1}{N_O} cN_W N_I + N_I}$$

Here, $c$ is the number of channels, $N_W$ is the width times the height of the filter, $N_I$ is the width times the height of the input of the layer, and $N_O$ is the number of XNOR operations that can be done in a single clock cycle. In [1], $c = 256$, $N_W = 14^2$, and $N_I = 3^2$ were used since layers with these parameters are common in ResNet [35]. While using a CPU, $N_O$ will be 64, which will be our baseline. Fig. 6(b) shows the speedup of our approach compared to XNOR-Net being executed in CPU. The speedup of this application compared to the CPU is significantly higher for our XNOR Implementation. We also compare our design with the existing works that require two or three cycles for XNOR operation. Additionally, our design scales better for larger array sizes than the existing designs. In addition to XNOR net, our design could also be used for XOR-Net [36], a version of XNOR-Net that uses XOR and reduces the required number of full precision operations significantly. Using this algorithm, we should see similar speedups and scaling as we did with XNOR-Net, though they will be slightly closer to the ideal $S = \frac{N_O}{64}$ speedup since XOR-Net reduces the full precision operations in a layer with our given parameters by 39.84% [36].

## VII.    Conclusion

In this paper, an all-CMOS single-cycle in-memory XOR/XNOR operation is proposed for bulk data copy, verification, and encryption making a slight modification in the peripheral circuitry. Our design allows for a reduced number of cycles and a leap in latency performance. For bulk data operations, even an incremental improvement can be tremendously advantageous. This circuit topology can also be used in modern and upcoming heavy data applications like binary convolutional neural networks. In-memory computing on top of that single-cycle operation can well justify the area implication and extra circuit overhead needed for the design.

## Acknowledgment

S. A. was supported with funds provided by the Science Alliance, a Tennessee Higher Education Commission center of excellence administered by The University of Tennessee-Oak Ridge Innovation Institute on behalf of The University of Tennessee, Knoxville.

N. S. was supported in part by the National Science Foundation under Grant No. 2132918.

This manuscript has also been authored in part by UT-Battelle, LLC under Contract No. DE-AC05-00OR22725 with the U.S. Department of Energy. The publisher, by accepting the article for publication, acknowledges that the U.S. Government retains a non-exclusive, paid-up, irrevocable, worldwide license to publish or reproduce the published form of the manuscript or allow others to do so, for U.S. Government purposes. The DOE will provide public access to these results in accordance with the DOE Public Access Plan (http://energy.gov/downloads/doe-public-access-plan).

## Data Availability

The data that support the plots within this paper and other findings of this study are available from the corresponding author upon reasonable request.

## Competing Interests

The authors declare no competing interests.

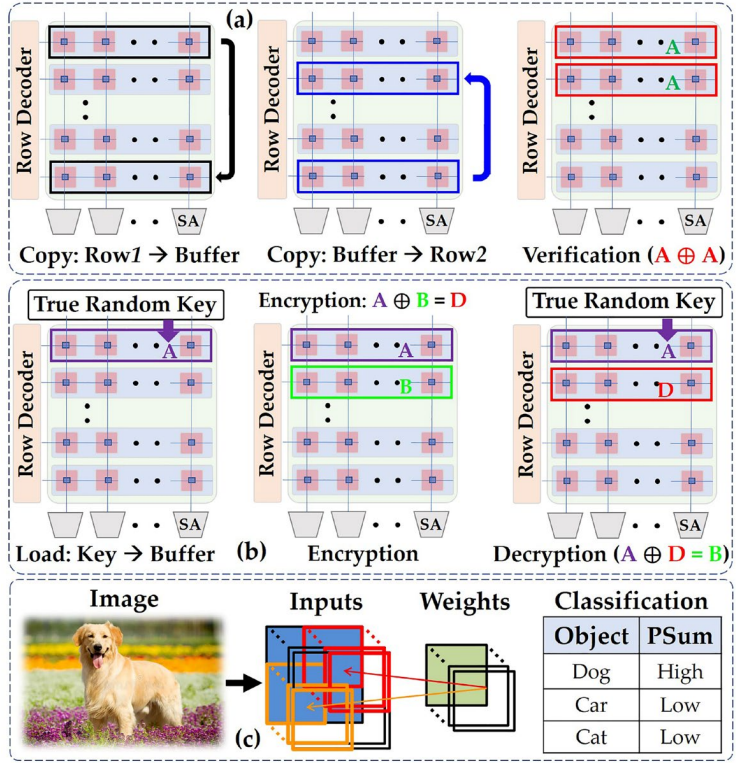

**FIGURE 1. A system level view in commercial memory products, where the memory cells are banked, will help understand the latency minimization for the proposed CiM XOR in (a) verification of copied data and (b) data encryption/decryption. (c) CiM configuration can also be used to deploy binary CNN to image classification problem which is essentially an XNOR operation.**

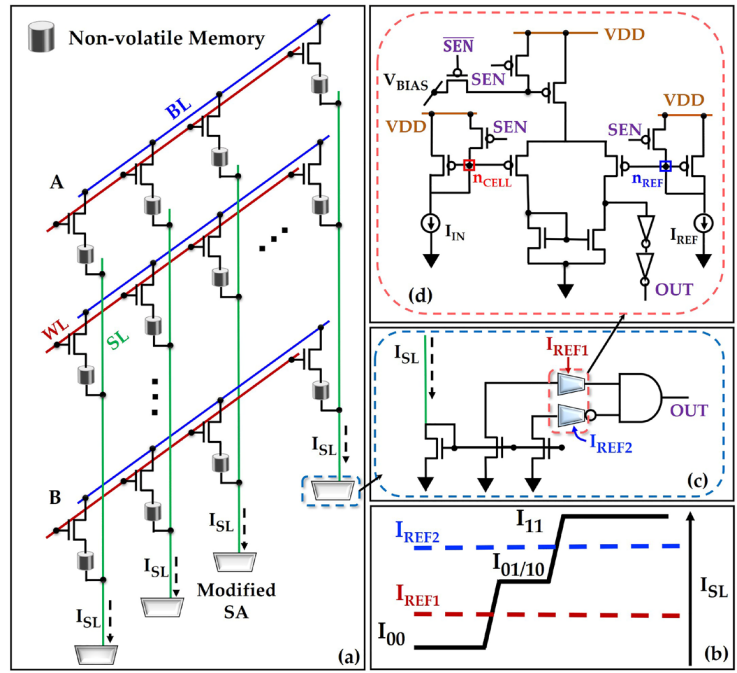

**FIGURE 2. (a) Non-volatile memory array with modified sense amplifiers. (b) Mechanism of choosing reference currents. Schematic of (c) the modified SA for in-memory XOR/XNOR and (d) a current sense amplifier.**

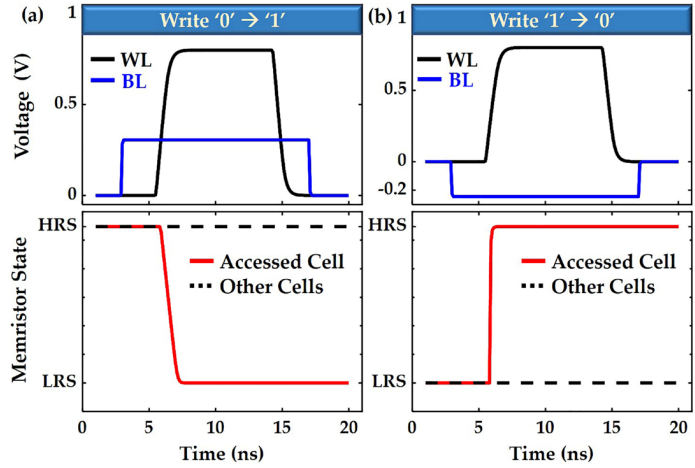

**FIGURE 3.** (a) Write '0' → '1' (HRS → LRS) and (b) '1' → '0' (LRS → HRS) operations upon applying suitable WL and BL biases.

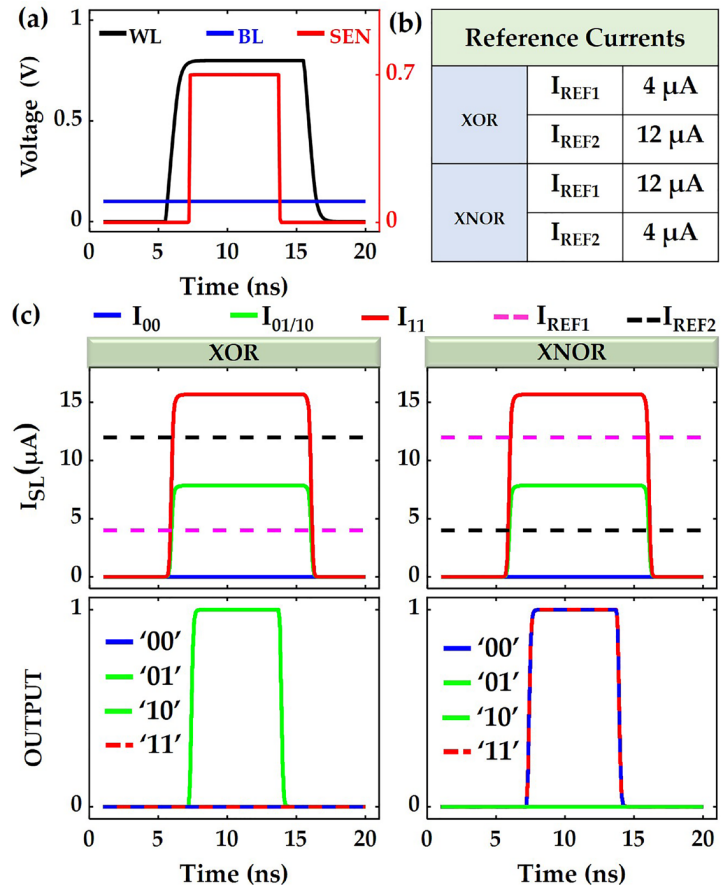

**FIGURE 4.** (a) The application of required voltages to WLs, BLs, and SEN. (b) Reference current levels chosen for XOR and XNOR operations. (d) SL currents and logic outputs of XOR and XNOR operations.

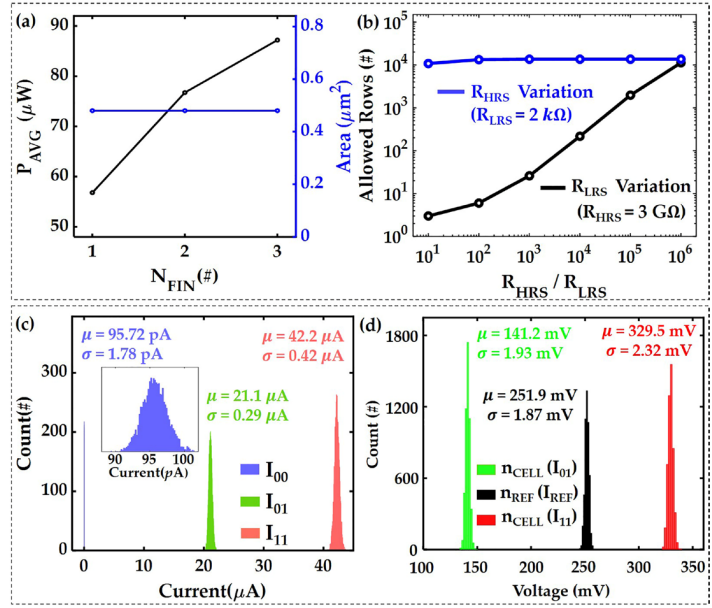

**FIGURE 5.** (a) Effect of number of fins of the transistors on the CSA circuit and (b) memristor on/off ratio on the array size. Histogram plots of (c) the current distributions and (d) voltages of $n_{CELL}$ and $n_{REF}$ nodes set by the distributions in input and reference current levels, respectively.

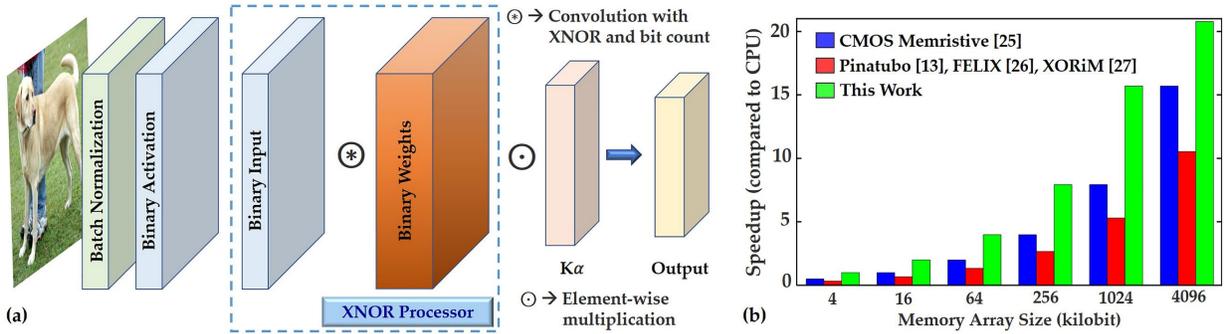

**FIGURE 6.** Comparison of our design with the existing works based on the implementation of a XNOR-based CNN.

**TABLE I: Comparison of our design with the existing works.**

| Design | Properties | | |
|---|---|---|---|
| | Tech. | Additional Transistors | Latency (Cycles) |
| Pinatubo [17] | CMOS | 7 | 3 |
| FELIX [31] | Crossbar | - | 3 |
| CMOS Memristive [30] | CMOS | 16 | 2 |
| XORiM [32] | CMOS | 12 | 3 |
| SiXOR [33] | Memristor | - | 1 |
| This Work | CMOS | 13 | 1 |

**References**

[1] B. H. McCarthy and S. Ponedal, "IBM Unveils World's First 2 Nanometer Chip Technology, Opening a New Frontier for Semiconductors," *IBM News Room*, pp. 6–8, 2021.

[2] S. Alam, M. M. Islam, M. S. Hossain, A. Jaiswal, and A. Aziz, "CryoCiM: Cryogenic compute-in-memory based on the quantum anomalous Hall effect," *Appl. Phys. Lett.*, vol. 120, no. 14, p. 144102, Apr. 2022, doi: 10.1063/5.0092169.

[3] V. Sze, Y. H. Chen, J. Einer, A. Suleiman, and Z. Zhang, "Hardware for machine learning: Challenges and opportunities," *Proc. Cust. Integr. Circuits Conf.*, vol. 2017-April, Jul. 2017, doi: 10.1109/CICC.2017.7993626.

[4] S. Kanev, J. P. Darago, K. Hazelwood, P. Ranganathan, T. Moseley, G.-Y. Wei, and D. Brooks, "Profiling a warehouse-scale computer," in *Proceedings of the 42nd Annual International Symposium on Computer Architecture*, 2015, pp. 158–169, doi: 10.1145/2749469.2750392.

[5] A. Boroumand, S. Ghose, Y. Kim, R. Ausavarungnirun, E. Shiu, R. Thakur, D. Kim, A. Kuusela, A. Knies, P. Ranganathan, and O. Mutlu, "Google Workloads for Consumer Devices: Mi-tigating Data Movement Bottlenecks," *Proc. Twenty-Third Int. Conf. Archit. Support Program. Lang. Oper. Syst.*, vol. 18, 2018, doi: 10.1145/3173162.

[6] S. Li, D. Niu, K. T. Malladi, H. Zheng, B. Brennan, and Y. Xie, "DRISA: A DRAM-based reconfigurable in-situ accelerator," *Proc. Annu. Int. Symp. Microarchitecture, MICRO*, vol. Part F1312, pp. 288–301, Oct. 2017, doi: 10.1145/3123939.3123977.

[7] Q. Deng, L. Jiang, Y. Zhang, M. Zhang, and J. Yang, "DrAcc: A DRAM based accelerator for accurate CNN inference," *Proc. - Des. Autom. Conf.*, vol. Part F1377, Jun. 2018, doi: 10.1145/3195970.3196029.

[8] S. Alam, J. Hutchins, M. S. Hossain, K. Ni, V. Narayanan, and A. Aziz, "Cryogenic In-Memory Matrix-Vector Multiplication using Ferroelectric Superconducting Quantum Interference Device (FE-SQUID)," in *2023 60th ACM/IEEE Design Automation Conference (DAC)*, 2023, pp. 1–6, doi: 10.1109/DAC56929.2023.10247669.

[9] Q. Dong, M. E. Sinangil, B. Erbagci, D. Sun, W. S. Khwa, H. J. Liao, Y. Wang, and J. Chang, "A 351TOPS/W and 372.4GOPS Compute-in-Memory SRAM Macro in 7nm FinFET CMOS for Machine-Learning Applications," *Dig. Tech. Pap. - IEEE Int. Solid-State Circuits Conf.*, vol. 2020-Febru, pp. 242–244, Feb. 2020, doi: 10.1109/ISSCC19947.2020.9062985.

[10] S. Alam, M. M. Islam, M. S. Hossain, A. Jaiswal, and A. Aziz, "Cryogenic In-Memory Bit-Serial Addition Using Quantum Anomalous Hall Effect-Based Majority Logic," *IEEE Access*, vol. 11, pp. 60717–60723, Jun. 2023, doi: 10.1109/ACCESS.2023.3285604.

[11] H. S. P. Wong, S. Raoux, S. Kim, J. Liang, J. P. Reifenberg, B. Rajendran, M. Asheghi, and K. E. Goodson, "Phase change memory," *Proc. IEEE*, vol. 98, no. 12, pp. 2201–2227, 2010, doi: 10.1109/JPROC.2010.2070050.

[12] "A 130.7-mm2 2-layer 32-gb reram memory device in 24-nm technology," *IEEE J. Solid-State Circuits*, vol. 49, no. 1, pp. 140–153, Jan. 2014, doi: 10.1109/JSSC.2013.2280296.

[13] "MRAM as Embedded Non-Volatile Memory Solution for 22FFL FinFET Technology," in *Technical Digest - International Electron Devices Meeting, IEDM*, 2019, vol. 2018-Decem, pp. 18.1.1-18.1.4, doi: 10.1109/IEDM.2018.8614620.

[14] D. Reis, S. Datta, M. T. Niemier, X. S. Hu, K. Ni, W. Chakraborty, X. Yin, M. Trentzsch, S. Dünkel, T. Melde, J. Müller, and S. Beyer, "Design and Analysis of an Ultra-Dense, Low-Leakage, and Fast FeFET-Based Random Access Memory Array," *IEEE J. Explor. Solid-State Comput. Devices Circuits*,

vol. 5, no. 2, pp. 103–112, Dec. 2019, doi: 10.1109/JXCDC.2019.2930284.

[15] S. Alam, M. S. Hossain, S. R. Srinivasa, and A. Aziz, "Cryogenic memory technologies," *Nat. Electron. 2023 63*, vol. 6, no. 3, pp. 185–198, Mar. 2023, doi: 10.1038/s41928-023-00930-2.

[16] D. Reis, M. Niemier, and X. Sharon Hu, "Computing in memory with FeFETs," *Proc. Int. Symp. Low Power Electron. Des.*, vol. 18, 2018, doi: 10.1145/3218603.

[17] S. Li, C. Xu, Q. Zou, J. Zhao, Y. Lu, and Y. Xie, "Pinatubo: A processing-in-memory architecture for bulk bitwise operations in emerging non-volatile memories," *Proc. - Des. Autom. Conf.*, vol. 05-09-June-2016, Jun. 2016, doi: 10.1145/2897937.2898064.

[18] W. Kang, H. Wang, Z. Wang, Y. Zhang, and W. Zhao, "In-Memory Processing Paradigm for Bitwise Logic Operations in STT-MRAM," *IEEE Trans. Magn.*, vol. 53, no. 11, Nov. 2017, doi: 10.1109/TMAG.2017.2703863.

[19] A. Govindankutty, S. Alam, S. Das, N. Challapalle, A. Aziz, and S. George, "Ternary In-Memory Computing with Cryogenic Quantum Anomalous Hall Effect Memories," in *Proceedings of the Great Lakes Symposium on VLSI*, 2023, pp. 521–526, doi: 10.1145/3583781.3590236.

[20] A. Govindankutty, S. Alam, S. Das, A. Aziz, and S. George, "Cryogenic In-memory Binary Multiplier Using Quantum Anomalous Hall Effect Memories," *Proc. - Int. Symp. Qual. Electron. Des. ISQED*, vol. 2023-April, 2023, doi: 10.1109/ISQED57927.2023.10129345.

[21] S. Alam, M. S. Hossain, and A. Aziz, "A non-volatile cryogenic random-access memory based on the quantum anomalous Hall effect," *Sci. Rep.*, 2021, doi: 10.1038/s41598-021-87056-7.

[22] S. Alam, M. M. Islam, M. S. Hossain, K. Ni, V. Narayanan, and A. Aziz, "Cryogenic Memory Array based on Ferroelectric SQUID and Heater Cryotron," *2022 Device Res. Conf.*, pp. 1–2, Jun. 2022, doi: 10.1109/DRC55272.2022.9855813.

[23] V. Seshadri, Y. Kim, C. Fallin, D. Lee, R. Ausavarungnirun, G. Pekhimenko, Y. Luo, O. Mutlu, P. B. Gibbons, M. A. Kozuch, and T. C. Mowry, "RowClone: Fast and energy-efficient in-DRAM bulk data copy and initialization," *MICRO 2013 - Proc. 46th Annu. IEEE/ACM Int. Symp. Microarchitecture*, pp. 185–197, 2013, doi: 10.1145/2540708.2540725.

[24] "Intel® 64 and IA-32 Architectures Software Developer Manuals." [Online]. Available: https://www.intel.com/content/www/us/en/developer/articles/technical/intel-sdm.html. [Accessed: 18-Jan-2022].

[25] V. Seshadri, K. Hsieh, A. Boroumand, D. Lee, M. A. Kozuch, O. Mutlu, P. B. Gibbons, and T. C. Mowry, "Fast Bulk Bitwise and and or in DRAM," *IEEE Comput. Archit. Lett.*, vol. 14, no. 2, pp. 127–131, Jul. 2015, doi: 10.1109/LCA.2015.2434872.

[26] X. Xin, Y. Zhang, and J. Yang, "Reducing DRAM access latency via helper rows," *Proc. - Des. Autom. Conf.*, vol. 2020-July, Jul. 2020, doi: 10.1109/DAC18072.2020.9218719.

[27] M. F. Chang, S. J. Shen, C. C. Liu, C. W. Wu, Y. F. Lin, Y. C. King, C. J. Lin, H. J. Liao, Y. Der Chih, and H. Yamauchi, "An offset-tolerant fast-random-read current-sampling-based sense amplifier for small-cell-current nonvolatile memory," *IEEE J. Solid-State Circuits*, vol. 48, no. 3, pp. 864–877, 2013, doi: 10.1109/JSSC.2012.2235013.

[28] N. Shukla, R. K. Ghosh, B. Gnsafe, and S. Datta, "Fundamental mechanism behind volatile and non-volatile switching in metallic conducting bridge RAM," *Tech. Dig. - Int. Electron Devices Meet. IEDM*, pp. 4.3.1-4.3.4, Jan. 2018, doi: 10.1109/IEDM.2017.8268325.

[29] "Arizona State University Predictive Technology Models." [Online]. Available: http://ptm.asu.edu/.

[30] S. Kvatinsky, A. Kolodny, U. C. Weiser, and E. G. Friedman, "Memristor-based IMPLY logic design procedure," *Proc. - IEEE Int. Conf. Comput. Des. VLSI Comput. Process.*, pp. 142–147, 2011, doi:

10.1109/ICCD.2011.6081389.

[31] S. Gupta, M. Imani, and T. Rosing, "FELIX: Fast and energy-efficient logic in memory," *IEEE/ACM Int. Conf. Comput. Des. Dig. Tech. Pap. ICCAD*, vol. 7, Nov. 2018, doi: 10.1145/3240765.3240811.

[32] K. Zou, Y. Wang, H. Li, and X. Li, "XORiM: A case of in-memory bit-comparator implementation and its performance implications," *Proc. Asia South Pacific Des. Autom. Conf. ASP-DAC*, vol. 2018-Janua, pp. 349–354, Feb. 2018, doi: 10.1109/ASPDAC.2018.8297348.

[33] N. Taherinejad, "SIXOR: Single-Cycle In-Memristor XOR," *IEEE Trans. Very Large Scale Integr. Syst.*, vol. 29, no. 5, pp. 925–935, May 2021, doi: 10.1109/TVLSI.2021.3062293.

[34] M. Rastegari, V. Ordonez, J. Redmon, and A. Farhadi, "XNOR-net: Imagenet classification using binary convolutional neural networks," in *European Conference on Computer Vision*, 2016, vol. ECCV 2016, pp. 525–542, doi: 10.1007/978-3-319-46493-0_32.

[35] K. He, X. Zhang, S. Ren, and J. Sun, "Deep residual learning for image recognition," *Proc. IEEE Comput. Soc. Conf. Comput. Vis. Pattern Recognit.*, vol. 2016-December, pp. 770–778, Dec. 2016, doi: 10.1109/CVPR.2016.90.

[36] S. Zhu, L. H. K. Duong, and W. Liu, "XOR-net: An efficient computation pipeline for binary neural network inference on edge devices," in *Proceedings of the International Conference on Parallel and Distributed Systems - ICPADS*, 2020, vol. 2020-Decem, pp. 124–131, doi: 10.1109/ICPADS51040.2020.00026.